\def\EE#1{\times 10^{#1}}
\def\ket#1{\left|#1\right\rangle}

\documentclass[aps,prl,twocolumn,floatfix]{revtex4}
\usepackage{graphicx,amsmath,amssymb,ulem}
\usepackage{color}
\begin{document}
\title{Parametric amplification of vacuum fluctuations in a spinor condensate}

\author{C.~Klempt$^1$, O.~Topic$^1$, G.~Gebreyesus$^2$, M.~Scherer$^1$, T.~Henninger$^1$, P.~Hyllus$^3$, W.~Ertmer$^1$, L.~Santos$^2$, J.J.~Arlt$^1$}

\affiliation{$^1$ Institut f\"ur Quantenoptik, Leibniz Universit\"at Hannover, Welfengarten~1, D-30167~Hannover}
\affiliation{$^2$ Institut f\"ur Theoretische Physik, Leibniz Universit\"at Hannover, Appelstra\ss e~2, D-30167~Hannover, Germany}
\affiliation{$^3$ BEC-INFM, Dipartimento di Fisica, Universit{\`a} di Trento, Via Sommarive 14, I-38050 Povo, Italy}

\date{\today}

\begin{abstract}
Parametric amplification of vacuum fluctuations is crucial
in modern quantum optics, enabling the creation of
squeezing and entanglement. We demonstrate the parametric amplification
of vacuum fluctuations for matter waves using a spinor $F=2$ $^{87}$Rb
condensate. Interatomic interactions
lead to correlated pair creation in the $m_F=\pm 1$ states from an initial
unstable $m_F=0$ condensate, which acts as a vacuum for $m_F\neq 0$.
Although this pair creation from a pure $m_F=0$ condensate
is ideally triggered by vacuum fluctuations, unavoidable spurious
initial $m_F=\pm 1$ atoms induce a classical seed which may
become the dominant triggering mechanism.
We show that pair creation is insensitive to a classical seed
for sufficiently large magnetic fields, demonstrating the dominant
role of vacuum fluctuations. The presented system thus provides
a direct path towards the generation of non-classical states of matter
on the basis of spinor condensates.
\end{abstract}

\maketitle
Parametric amplification of vacuum fluctuations in nonlinear media~\cite{Walls1994}
plays a crucial role in the investigation and application of non-classical states of light. These states have revolutionized the field of quantum optics in the past decades. Since the first observation of squeezed light~\cite{Slusher1985}, these non-classical states of light have become a valuable tool in modern optics, e.g. for the enhancement of modern interferometers~\cite{Vahlbruch2005}. Similarly, the production of entangled photon pairs~\cite{Shih1988} has triggered an ongoing series of fundamental tests of modern quantum mechanics~\cite{Aspect1982,Weihs1998} and has many possible applications for quantum computing~\cite{Pan2001}. The tools developed for the production and manipulation of ultracold neutral atoms now bring many of these seminal investigations within the scope of experiments with matter waves. In this sense, the production of number-squeezed Bose-Einstein condensates (BECs)~\cite{Esteve2008} and spin squeezed thermal clouds~\cite{Hald1999} has been recently demonstrated.

Spinor BECs, consisting of atoms with non-zero spin $F$, provide an optimal non-linear medium for the production of non-classical states of matter. In these systems, inter-particle interactions lead to a coherent population transfer between different Zeeman $m_F$ sublevels (spin dynamics). The case of a sample prepared in $m_F=0$ (represented by $\ket{0}$) is particularly interesting. In that case the initial stages of the spin dynamics are characterized by the production of correlated atom pairs in $\ket{\pm 1}$~\cite{Duan2000,Pu2000}, resembling the production of Einstein-Podolsky-Rosen (EPR) pairs in optical parametric down conversion~\cite{Aspect1982}.

Ideally a pure initial $\ket{0}$ BEC acts as a vacuum for atoms in $m_F\neq 0$. Hence pair creation into $\ket{\pm 1}$ can be understood as a parametric amplification of quantum vacuum fluctuations. However, parametric amplifiers are exponentially sensitive to any spurious initial seed in the amplified modes, which may easily dominate the effect of vacuum fluctuations. Despite careful purification procedures a tiny number of spurious atoms in $\ket{\pm 1}$ is experimentally unavoidable. Hence it is crucial to determine whether these spurious seed atoms play a significant role in the spin dynamics. Only in the case they do not, the system may safely be considered a parametric amplifier for vacuum fluctuations.

\begin{figure}
\centering
\includegraphics*[width=7.0cm]{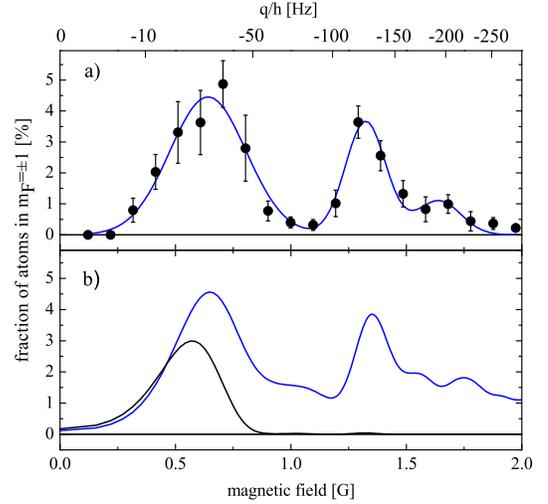}
\caption{(a) Fraction of atoms transferred into $\ket{m_F=\pm1}$ within $21$~ms as a function of the applied magnetic field~\cite{Klempt2009}. The error bars indicate statistical uncertainties. The blue line is a triple Gaussian fit to guide the eye. (b) Theoretical prediction for our experimental parameters (see text), with an initial BEC in $\ket{0}$ with $N=50000$ and $N_s=2$ seed atoms. The black line shows the GP result, whereas the blue line shows the result of our calculation, including both this classical seed and vacuum spin fluctuations.}
\label{fig:resonance}
\end{figure}

In this Letter we investigate the triggering mechanism for the parametric amplification of correlated pairs. Similar to the gain of the amplification, studied in Ref.~\cite{Klempt2009}, the triggering mechanism and its sensitivity depend crucially on the interplay between Zeeman energy, interactions and external confinement. In particular, the relative importance of triggering by a classical seed and by vacuum fluctuations depends on the magnetic field $B$. At low $B$ the amplification is strikingly sensitive to any unavoidable spurious seed. On the contrary for sufficiently large $B$ the classical seed plays no significant role, and the system can indeed be characterized as a parametric amplifier of vacuum fluctuations. The amplification of vacuum fluctuations opens fascinating perspectives towards the analysis of two-mode squeezing and EPR entanglement of matter waves on the basis of unstable spinor BECs~\cite{Duan2000,Pu2000}.

The onset of the spin dynamics is characterized by pair creation into
$\ket{\pm 1}$. In order to analyze this initial
regime we consider the perturbed spinor
$\hat{{\bf \Psi}}(\vec r,t)$ $=({\mathbf \Psi}_0 (\vec r)+\delta\hat{\mathbf \Psi}(\vec r,t))e^{-i\mu t}$, where
${\mathbf \Psi}_0(\vec r)=(0, 0, n_0(\vec r)^{1/2}, 0, 0)^T$
 represents the initial BEC in  $\ket{0}$, $\mu$ is the chemical potential, and $\delta\hat{{\mathbf \Psi}}(\vec r,t)= (\delta\hat\psi_{-2},\delta\hat\psi_{-1},\delta\hat\psi_{0},\delta\hat\psi_{1},\delta\hat\psi_{2})^T$
describes fluctuations. Up to second order in $\delta\hat{\mathbf \phi}_{\pm}=(\delta\hat{\mathbf \psi}_{+1}\pm \delta\hat{\mathbf \psi}_{-1})/\sqrt{2}$, pair creation is described by the Hamiltonian $\hat H=\hat H_+ + \hat H_-$, where
\begin{equation}
\hat H_{\pm}=\int d^3 r
\left [ \delta\hat\phi_{\pm}^\dag
\left( \hat H_{eff} +q \right ) \delta\hat\phi_{\pm}
\pm\frac{\Omega}{2}
\left ( \delta\hat\phi_{\pm}^2 +h.c \right ) \right ].
\label{eq:H}
\end{equation}
Here, $q\propto B^2$ characterizes the quadratic Zeeman energy (QZE),
$\hat H_{eff}=-\hbar^2\nabla^2/2m+V_{trap}(\vec r)+(U_1+U_0)n_0(\vec r)-\mu$,
$V_{trap}(\vec r)$ is the harmonic trap, $\Omega(\vec r)=U_1n_0(\vec r)$, and
$U_0= (7g_0+10g_2+18g_4)/35$ and $U_1= (-7g_0-5g_2+12g_4)/35$ characterize
the spin-preserving and spin-changing collisions. The coupling constant associated with the $s$-wave collisional channel with total spin $F$ is $g_F=4\pi\hbar^2a_F/m$.
Note that $\hat H_\pm$ is identical to the Hamiltonian describing an optical parametric amplifier~\cite{Meystre2007}.

The problem is analyzed best in the
basis of eigenstates $\hat H_{eff}\varphi_n(\vec r)=\epsilon_n\varphi_n(\vec r)$. By introducing
$\delta\hat\phi_{\pm}=\sum_n \hat b_{n,\pm}\varphi_n$ and
$\Omega_{n,n'}=\int d^3r \Omega\varphi_n\varphi_{n'}$ we may rewrite the Hamiltonian
\begin{equation}
\hat H_\pm=\sum_{n} \left ( \epsilon_n+q \right ) \hat b_{n,\pm}^\dag\hat b_{n,\pm}
\pm\sum_{n,n'}\frac{\Omega_{n,n'}}{2} \left ( \hat b_{n,\pm}\hat b_{n',\pm} + h.c  \right ).
\end{equation}
The Heisenberg equation $i\hbar\frac{d}{dt}\hat O_\nu^\pm=\left[H_\pm, \hat O_\nu^\pm\right]=\xi_\nu^\pm \hat O_\nu^\pm$ then yields the eigenvalues $\xi_\nu^\pm$
and the corresponding eigenoperators $\hat O_\nu^\pm=\sum_{n=1}^M \left ( R_{\nu,n}^\pm b_{n,\pm}+R_{\nu,M+n}^\pm b_{n,\pm}^\dag \right )$, where $M$
indicates the maximal $\varphi_n$ level considered.

Imaginary eigenvalues with Im$(\xi_\nu)>0$ result in an exponential amplification of
spin fluctuations, which indicates the onset of pair creation. In that case, the amplification dynamics is dominated by the most unstable mode $\nu_0$ with the largest imaginary part Im$(\xi_{\nu_0})=\hbar\Lambda(q)$. The instability rate $\Lambda(q)$ generally shows a non-monotonous multi-resonant $q$-dependence due to the interplay
between QZE, interactions and external confinement~\cite{Klempt2009}.

Based on this diagonalization of the Hamiltonian we can evaluate the experimentally relevant quantum evolution of the operators $a_{n,m_F}$ and $a_{n,m_F}^\dag$ where $a_{n,\pm1}=(b_{n,+}\pm b_{n,-})/\sqrt{2}$.
The total population in
$\ket{\pm 1}$ is then $P_{m_F}(t)=\sum_n \langle a_{n,m_F}^\dag(t) a_{n,m_F}(t) \rangle$, where the average
is performed over the initial state $\ket{\Psi_{m_F}(0)}=\ket{\Psi}$.

In the presence of unstable spin excitation modes, the initial state $\ket{\Psi}$ triggers the subsequent
amplification. Ideally $\ket{\Psi}$ should be a state
with no particles in $m_F=\pm 1$ (defining our vacuum state $\ket{vac}$). In that case, pair creation is triggered by vacuum fluctuations (quantum triggering). However, unavoidable slight
imperfections always lead to $N_s\ll N$ spurious atoms in $\ket{\pm 1}$ at $t=0$. Although small, these initial impurities serve as a classical seed of the spin dynamics (classical triggering), which can dominate the quantum triggering. It is hence crucial to determine the relative role of quantum and classical triggering.

Note that, in our experiments ~\cite{Klempt2009} the BEC in the $\ket{0}$ state is purified by quickly removing all atoms in the $\ket{\pm 1}$ states. This results in a non-equilibrium state and we stress that the time after purification is too short to produce any thermal population of the $\ket{\pm1}$ states. Thus, thermal spin fluctuations do not play any role in our discussion and  impurities in $\ket{\pm 1}$ can only be created by radio-frequency noise and magnetic field jitter after the purification. Since these processes act equally on each single atom, any spurious $\ket{\pm 1}$ atom is produced in the same spatial mode as the original BEC.
These spurious single-atom processes may also transfer a
tiny fraction of the thermal cloud into $\ket{\pm 1}$. However these atoms
lack significant spatial overlap with the most unstable excitations, and hence
do not contribute to the spin dynamics.

To evaluate the triggering mechanism, the initial state $\ket{\Psi}$ including a classical seed can be represented by $\ket{\Psi} = (N_s!)^{-1} (\hat\Phi_{+1}^\dag \hat\Phi_{-1}^\dag )^{N_s}\ket{{\rm vac}}$, where $\hat\Phi_{m_F}^\dag$ ($\hat\Phi_{m_F}$)  creates (annihilates) a $m_F$ particle
in the mode of the initial BEC. In the basis of the effective Hamiltonian, they can be represented by a vector $\mathbf{\Phi}$ where $\hat\Phi_{m_F}=\sum_n \Phi_n\hat a_{n,m_F}$, with $\Phi_n=\int d^3 r \sqrt{n_0}\varphi_n$. Now, one obtains the total population as a sum of a classically and a quantum triggered contribution $P_{m_F}(t)=P_C(t)+P_Q(t)$, where

\begin{eqnarray}
P_C(t)&=&N_s {\bf \Phi}^* \cdot \left (
\hat U^\dag(t)\cdot\hat U(t)+\hat V(t)^\dag\cdot\hat V(t)
\right )\cdot {\bf \Phi},  \label{eq:PC}\\
P_Q(t)&=&Tr\left\{\hat V^\dag(t)\cdot\hat V(t)\right\} \label{eq:PQ}.
\end{eqnarray}
and the operators $\hat U(t)$ and $\hat V(t)$ are the time evolution operators according to $\hat a_{n,m_F}(t)=\sum_{n'} \left [ U_{n,n'}(t)\hat a_{n',m_F}(0)+V_{n,n'}(t)\hat a_{n',m_F}^\dag(0)\right ]$.

We have employed Eqs.~(\ref{eq:PC}) and (\ref{eq:PQ}) to determine the population at any time $t$.
Note however that $F=2$ $^{87}$Rb BECs present inherent hyperfine-changing losses~\cite{Schmaljohann2004},
with an experimentally determined loss rate $\Gamma\simeq 10^{-2}$ ms$^{-1}$. Although these losses are small ($<20\%$ of the total number) during the typical investigation times of $20$ms, they may significantly alter the spin dynamics, mainly due to the dynamical modification of the resonant conditions for the instability
rate $\Lambda(q)$. We have carefully included these losses by splitting the evolution time into sub-intervals. At the beginning of each sub-interval we introduce losses and recalculate the evolution operator (due to the moderate loss rate typically few
sub-intervals already lead to convergence). Figure~\ref{fig:resonance} (b) shows the numerical result
for our experimental parameters. Note that the multi-resonant $q$ dependence of the instability rate $\Lambda(q)$ discussed in Ref.~\cite{Klempt2009} directly maps into a multi-peaked pair creation efficiency,
which is in very good agreement with our previous experimental results~\cite{Klempt2009} shown in
Fig.~\ref{fig:resonance} (a). As we discuss in detail below, the unknown average classical seed $N_s$ only
influences the spin dynamics significantly at the low field resonance, and hence it is obtained from a fit to
the relative fraction of atoms on the two resonances. 
Our theoretical results for the amplification dynamics depend
on the total number of atoms and especially on the precise values of the
scattering lengths $a_F$ \footnote{We use a$_0=87.9(2) a_B$, a$_2=91.2(2) a_B$, a$_4=99.0(2) a_B$ for the scattering lengths (in units of the Bohr radius $a_B$) which where obtained from a coupled channel analysis in the group of E. Tiemann.}. We varied these parameters within
their rather strict uncertainties (one standard deviation) obtaining a
very good agreement with our experimental results (Fig.~\ref{fig:resonance}).

Figure~\ref{fig:resonance} (b) also shows that the spin dynamics differs significantly from the result of a simple mean-field Gross-Pitaevskii (GP) approach in which field operators $\hat\psi_{m_F}({\vec{r}})$ are substituted by c-number fields $\psi_{m_F}({\vec{r}})$. Quantum fluctuations are absent in this description and the spin dynamics can only be triggered classically with $\psi_{\pm 1}({\vec{r}})=\epsilon\psi_0({\vec{r}})$, with $\epsilon\ll 1$ (for more sophisticated approaches, see e.g. ref.~\cite{Saito2007}). The striking difference between GP and exact results indicates that the two resonances display a very different sensitivity to a classical seed and vacuum fluctuations.

\begin{figure}
\centering
\includegraphics*[width=7cm]{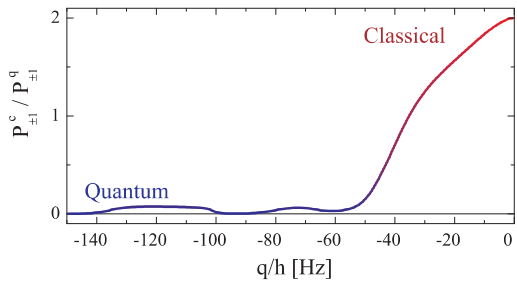}
\caption{Ratio between the classically-triggered $P^c$ and quantum-triggered $P^q$ populations for our experimental parameters, considering only the growth due to the most unstable mode. The quantum triggered dynamics is characterized by $P^c \ll P^q$.}
\label{fig:ratio}
\end{figure}

We estimate the relative importance of both triggering mechanisms by considering only the dominant contribution by the most unstable mode. In that case we may approximate the evolution of $P_{C,Q}(t)\simeq{\bar P}_{C,Q}\exp(2\Lambda(q) t)$ and evaluate the time-independent ratio $\eta={\bar P}_C/{\bar P}_Q$ between the classical and quantum triggering mechanisms, shown in Fig.~\ref{fig:ratio}. For sufficiently small $|q|$, including the low field resonance shown in Fig.~\ref{fig:resonance}, any classical seed is highly relevant ($\eta\sim 1$) due to the large overlap between the wavefunctions of the most unstable mode and the original BEC. However, for larger $|q|$, including the high field resonance, the population triggered by quantum vacuum fluctuations dominates ($\eta\ll 1$)
\footnote{An equivalent analysis shows that recent amplification experiments~\cite{Leslie2009}
could be dominantly quantum-triggered only for $q/h \lesssim 6$~Hz.}.
As detailed below, our experiments confirm this remarkable difference in sensitivity of the two resonances to any spurious classical seed. Note that this approach to determine the triggering mechanism avoids the difficulties arising in e.g. the investigation of fluctuations of the amplifier output~\cite{Leslie2009}, which is largely impeded by the significant statistical uncertainty of the atom number measurement (detection noise) and small shot-to-shot variations of the total atom number which lead to fluctuations in the pair creation efficiency (amplifier noise).

To initiate our experiments, we prepare BECs containing $7\EE{4}$  $F=2$ $^{87}$Rb atoms in the  $\ket{0}$ state in an optical dipole trap with trapping frequencies of $(176,132,46)$~Hz. We carefully remove residual atoms in other spin components by briefly applying a strong magnetic field gradient of $\approx 50$~G/cm, which is ramped down within $10$~ms. To investigate the spin dynamics, the applied homogeneous magnetic field is subsequently lowered from $7.9$~G to a specific value between $0.12$ and $2$~G within $3$~ms. The BEC is held at the chosen magnetic field for a variable time to allow for spin changing collisions. Finally, the number of atoms in all $m_F$ components is measured independently by applying a strong magnetic field gradient during time-of-flight expansion. To evaluate the onset of the exponential amplification regime, we restrict our investigation to short spin evolution times and small populations in $\ket{\pm 1}$.

\begin{figure}
\includegraphics*[width=7.0cm]{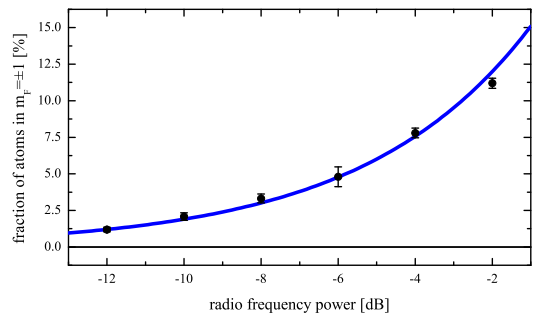}
\caption{Fraction of produced seed atoms in $\ket{m_F=\pm 1}$ as a function of the radio frequency power on a logarithmic scale. The blue line indicates a linear fit to the data which allows for an extrapolation to very few seed atoms.}
\label{fig3}
\end{figure}

The sensitivity of the system to a classical seed is investigated by deliberately producing a very small symmetric seed population in the $\ket{\pm 1}$ states prior to the spin evolution. This is accomplished by using a radio frequency pulse which transfers a variable number of atoms from the $\ket{0}$ to $\ket{\pm 1}$ states. This pulse is applied for $5\ \mu$s at a magnetic field of $7.9$~G. A frequency of $5.6$~MHz was identified, which couples the BEC in the $\ket{0}$ state symmetrically to the  $\ket{\pm 1}$ states. The number of transferred atoms was calibrated at the smallest detectable numbers in the linear regime. Figure \ref{fig3} shows a linear fit to the data which allows for an extrapolation to very small atom numbers. By reducing the radio-frequency power by a further 15 to $25$~dB, it is thus possible to reproducibly prepare a very small number of seed atoms.
Note that both spuriously and deliberately produced seeds result from similar single-atom processes, and hence have exactly the same spatial dependence as the original BEC. The sensitivity to the deliberately produced seed is therefore representative of the sensitivity to any spurious seed in the experiment.

\begin{figure}[ht]
\centering
\includegraphics*[width=7.0cm]{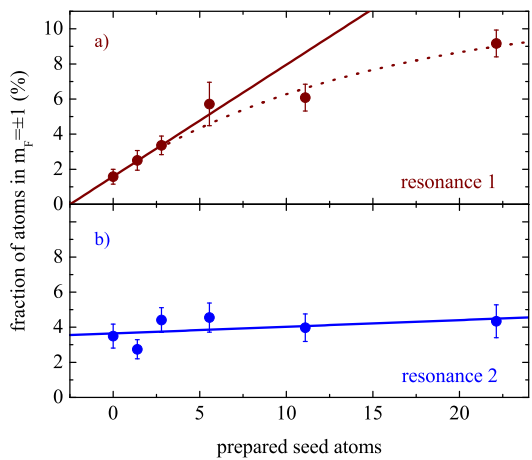}
\caption{Fraction of atoms transferred into $\ket{m_F=\pm 1}$ as a function of the deliberately prepared number of seed atoms. (a) The fraction recorded at $0.65$~G, corresponding to the low field resonance, shows a strong dependence on the classical seed after an evolution time of $15$~ms. (b) The fraction recorded after $t=23$~ms on the high field resonance at $1.29$~G is independent of the number of seed atoms and we conclude that it is triggered by vacuum spin fluctuations. The error bars indicate statistical uncertainties. The solid lines are fits and the dotted line is a guide to the eye indicating saturation.}
\label{fig:seed}
\end{figure}

Figure~\ref{fig:seed} (a) shows the fraction of atoms in the $\ket{\pm 1}$ states after the spin dynamics depending on the number of seed atoms  for the low field resonance (Fig.~\ref{fig:resonance}). Starting at a small offset value, this fraction grows linearly with increasing number of seed atoms (amplification of $23$~dB) and starts to saturate for transferred fractions above $6\%$. Hence the low field resonance is strikingly sensitive to a classical seed, down to an extremely small number of seed atoms. The fact that the spin dynamics is sensitive to very few seed atoms also proves that the seed production works reliably and is not dominated by radio-frequency background noise. The offset is both due to the amplification of vacuum fluctuations and a small number of accidentally produced seed atoms.


However this sensitivity to a classical seed is not general. Figure~\ref{fig:seed} (b) shows the fraction of transferred atoms  for the high field resonance.  For this resonance the fraction of $\ket{\pm 1}$ atoms is independent of the number of deliberately produced seed atoms, as expected from Fig.~\ref{fig:ratio}, which indicates that a spurious classical seed is irrelevant for sufficiently large $|q|$. This experiment thus confirms that classical seed cannot trigger the spin dynamics in this mode and we conclude that pair creation indeed acts as a parametric
amplification of vacuum fluctuations on the high field resonance.

In conclusion, we have shown that a spinor $F=2$ BEC initially prepared
in an unstable $\ket{0}$ state can provide a parametric amplifier for
vacuum fluctuations, where the $\ket{0}$ condensate acts as
an effective vacuum for atoms in $\ket{\pm 1}$. Similar to other parametric amplifiers
the system is exponentially sensitive to a spurious classical seed.
We have therefore carefully analyzed the dependence of the pair creation efficiency on a classical seed in $\ket{\pm 1}$. For low magnetic fields
the amplification is extremely sensitive to spurious classical seed atoms, whereas for large enough fields the classical seed is irrelevant, and the observed pair creation is due to a parametric amplification of vacuum fluctuations. This mechanism is identical to spontaneous optical parametric down conversion and paves the way for the development of non-classical atom optics on the basis of unstable spinor condensates. In particular, this system provides a direct path towards the observation of two-mode squeezing of matter waves, and a promising method for the creation of entangled atomic EPR pairs~\cite{Duan2000,Pu2000}. In addition magnetic dipole-dipole interactions may play a significant role~\cite{Vengalattore2008,Swislocki2009} for the case of $F=1$ BECs and will be the subject of future investigations.

We thank E. Tiemann and F. Deuretzbacher for fruitful discussions. We acknowledge support from the Centre for Quantum Engineering and Space-Time Research QUEST, from the Deutsche Forschungsgemeinschaft (SFB 407), and the European Science Foundation (EuroQUASAR).

\bibliography{bibliothek}

\begin{thebibliography}{10}

\bibitem{Walls1994}
D.~F. Walls and G. Milburn, {\em Quantum Optics} (Springer, Heidelberg, 1994).

\bibitem{Slusher1985}
R.~E. Slusher {\it et~al.}, Phys. Rev. Lett. {\bf 55},  2409  (1985).

\bibitem{Vahlbruch2005}
H. Vahlbruch {\it et~al.}, Phys. Rev. Lett. {\bf 95},  211102  (2005).

\bibitem{Shih1988}
Y.~H. Shih and C.~O. Alley, Phys. Rev. Lett. {\bf 61},  2921  (1988).

\bibitem{Aspect1982}
A. Aspect, J. Dalibard, and G. Roger, Phys. Rev. Lett. {\bf 49},  1804  (1982).

\bibitem{Weihs1998}
G. Weihs {\it et~al.}, Phys. Rev. Lett. {\bf 81},  5039  (1998).

\bibitem{Pan2001}
J.-W. Pan {\it et~al.}, Phys. Rev. Lett. {\bf 86},  4435  (2001).

\bibitem{Esteve2008}
J. Esteve {\it et~al.}, Nature {\bf 455},  1216  (2008).

\bibitem{Hald1999}
J. Hald, J.~L. S\o{}rensen, C. Schori, and E.~S. Polzik, Phys. Rev. Lett. {\bf
  83},  1319  (1999).

\bibitem{Duan2000}
L.-M. Duan, A. S\o{}rensen, J.~I. Cirac, and P. Zoller, Phys. Rev. Lett. {\bf
  85},  3991  (2000).

\bibitem{Pu2000}
H. Pu and P. Meystre, Phys. Rev. Lett. {\bf 85},  3987  (2000).

\bibitem{Klempt2009}
C. Klempt {\it et~al.}, arXiv:0902.2058  (2009).

\bibitem{Meystre2007}
P. Meystre and M. Sargent, {\em Elements of quantum optics}, 4th ed. (Springer,
  Berlin, 2007).

\bibitem{Schmaljohann2004}
H. Schmaljohann {\it et~al.}, Phys. Rev. Lett. {\bf 92},  040402  (2004).

\bibitem{Saito2007}
H. Saito, Y. Kawaguchi, and M. Ueda, Phys. Rev. A {\bf 76},  043613  (2007).

\bibitem{Leslie2009}
S.~R. Leslie {\it et~al.}, Physical Review A (Atomic, Molecular, and Optical
  Physics) {\bf 79},  043631  (2009).

\bibitem{Vengalattore2008}
M. Vengalattore, S.~R. Leslie, J. Guzman, and D.~M. Stamper-Kurn, Phys. Rev.
  Lett. {\bf 100},  170403  (2008).

\bibitem{Swislocki2009}
T. \'Swis\l{}ocki, M. Brewczyk, M. Gajda, and K. Rzk{a}\.zewski,
  arXiv:0901.1763  (2009).

\end{thebibliography}

\end{document}